# Novel Outer Bounds and Capacity Results for the Interference Channel with Conferencing Receivers

Reza K. Farsani, Amir K. Khandani
Department of Electrical and Computer Engineering
University of Waterloo
Waterloo, ON., Canada
Email: {r3khosra, khandani}@uwaterloo.ca

*Abstract*—In this paper, capacity bounds for the two-user Interference Channel (IC) with cooperative receivers via conferencing links of finite capacities is investigated. The capacity results known for this communication scenario are limited to a very few special cases of the one-sided IC. One of the major challenges in analyzing such cooperative networks is how to establish efficient capacity outer bounds for them. In this paper, by applying new techniques, novel capacity outer bounds are presented for the ICs with conferencing users. Using the outer bounds, several new capacity results are proved for interesting channels with unidirectional cooperation in strong and mixed interference regimes. A fact is that the conferencing link (between receivers) may be utilized to provide one receiver with information about its corresponding signal or its non-corresponding signal (interference signal). As a remarkable consequence, it is demonstrated that both strategies can be helpful to achieve the capacity of the channel. Finally, for the case of Gaussian IC, it is mathematically shown that our outer bound is strictly tighter than the previous one derived by Wang and Tse.

## I. INTRODUCTION

One of the main challenges to establish a reliable wireless communication network with a satisfactory performance is how to manage the interference effect caused by concurrent signaling of different users. In network information theory, this key aspect of wireless communication systems is basically addressed by the Interference Channel (IC). Another important feature of many practical wireless systems is the feasibility of cooperation among different users that allows them to exchange information. User cooperation has been shown to be a crucial way of improving performance of communication networks [1]. Specifically, it is an effective way to mitigate the interference in networks [2].

One of significant ways to set up cooperation among transmitters/receivers in a communication network is the use of conferencing links of finite capacities. In particular, modern cellular systems typically rely on some high capacity direct link between base-stations. Such configurations fall under the umbrella of channels with conferencing transmitters and/or conferencing receivers. Cooperation via conferencing links was first studied by Willems [3] for a Multiple Access Channel (MAC). Willems characterized the capacity region of the two-user MAC with conferencing transmitters. In the past decade, various communication scenarios with conferencing transmitters/receivers have been studied in network information theory [4]–[15]. In this paper, we study the two-user IC with conferencing receivers. This means, two clients, forming a two-user IC, send data to their respective base-stations, and the two base-stations are connected through links of given capacities. This scenario has been previously considered in several papers. Specifically, the capacity region of the Gaussian IC with conferencing receivers was established in [8] to within a constant gap. Other works in this regard include [10-15]. Despite considerable work on the cooperative interference channels with conferencing links, up to our knowledge, the available capacity results are limited to a very few special cases of the one-sided IC with unidirectional conferencing between receivers [14, 15]. In fact, the capacity of the two-user fully-connected IC with conferencing users was not previously known even for any of the special cases where the capacity is known for the IC without cooperation, for example, the strong interference channel [16].

Indeed, a major challenge to analyze cooperative networks in general and the IC with conferencing users in specific is how to establish efficient capacity outer bounds. In literature, generally there exist three types of outer bounds for cooperative interference networks: cut-set bounds [17], Sato type outer bounds [18], [19], and genie-aided outer bounds [8, 9]. The cut-set bounds and Sato type outer bounds are usually insufficient to derive capacity results or even to establish capacity to within constant-gap results for Gaussian channels. The outer bounds derived by genie-aided techniques, similar to [8, 9], are useful to establish constant-gap results for Gaussian channels; however, they still need to be tightened to derive exact capacity results. In this paper, we present a novel outer bound for the two-user IC with conferencing receivers. The derivation of our outer bound is indeed involved in subtle applications of the Csiszar-Korner identity [20] for manipulating multi-letter mutual information functions to establish consistent and well-formed single-letter constraints on the communication rates. In fact, we derive our bound by extending the constraints of the outer bound established in [21-22] for the IC without cooperation (which was shown to be useful to derive several capacity results) to the conferencing settings as well as presenting constraints with new structures. Using our outer bound, we prove several new capacity results for the two-user (fully-connected) IC with conferencing users for both discrete and Gaussian cases. In particular, we derive four capacity results for interesting channels with unidirectional cooperation in mixed and strong interference regimes. It is a fact that a conferencing link (between receivers) may be utilized to provide one receiver with information about its corresponding signal or its non-corresponding signal (interference). As a remarkable consequence, we demonstrate that both strategies can be helpful to achieve capacity for the IC with conferencing receivers.

Finally, for the case of Gaussian IC, we show that the derived outer bound can be tightened more by introducing additional constraints which are derived by utilizing genie-aided techniques as well. As a result, we obtain a new outer bound for the Gaussian IC with conferencing receivers, which can be mathematically shown that is strictly tighter than the previous one obtained by Wang and Tse [9].

The rest of the paper is organized as follows. In Section II, channel models and definitions are given. The main results are presented in Section III and the paper is concluded in Section IV. Due to limited space, some of the proofs are given in the Appendix.

## II. CHANNEL MODELS

In this paper, a Random Variable (RV) is denoted by an upper case letter (e.g., $X$) and a lower case letter is used to show its realization (e.g., $x$). The alphabet of $X$ is represented by $\mathcal{X}$. The notation $[1:N]$ represents the set of integers from 1 to $N$, and $\|\mathcal{S}\|$ denotes the cardinality of the set $\mathcal{S}$.

**Definition 1.** The two-user IC is a communication scenario where two transmitters send independent messages to their corresponding users via a common media. The channel is given by the input signals $X_1 \in \mathcal{X}_1$ and $X_2 \in \mathcal{X}_2$, the outputs $Y_1 \in \mathcal{Y}_1$ and $Y_2 \in \mathcal{Y}_2$ and the transition probability function $\mathbb{P}(y_1, y_2 | x_1, x_2)$. The Gaussian channel is usually formulated in the following standard form:

$$Y_1 = S_{11} X_1 + S_{12} X_2 + Z_1$$
$$Y_2 = S_{21} X_1 + S_{22} X_2 + Z_2$$
(1)

where $Z_1$ and $Z_2$ are zero-mean unit-variance Gaussian RVs and $\mathbb{E}[X_i^2] \leq P_i, i = 1,2$.

*Conferencing Decoders:* The two-user IC with conferencing decoders is depicted in Fig. 1. For this channel, a length-$n$ code with $L_d$ conferencing rounds, denoted by $\mathbb{C}^n(L_d, R_1, R_2, D_{12}, D_{21})$ is described as follows.

The message $M_i$, which is uniformly distributed over the set $\mathcal{M}_i = [1:2^{nR_i}]$, is transmitted by the $i^{th}$ transmitter and decoded by the $i^{th}$ receiver, $i = 1,2$. The code includes two encoder functions as:

$$\nabla_i : \mathcal{M}_i \to \mathcal{X}_i^n, \qquad X^n = \nabla_i(M_i), \qquad i \in \{1,2\}$$

Each transmitter encodes its message by the respective encoding function and sends the generated codeword over the channel. The receiver $Y_i$ receives a sequence $Y_i^n \in \mathcal{Y}_i^n$. Before decoding process, the decoders hold a conference. The code consists of two sets of conferencing functions $\{\vartheta_{12,l}\}_{l=1}^{L_d}$ and $\{\vartheta_{21,l}\}_{l=1}^{L_d}$ with the corresponding output alphabets $\{\mathcal{V}_{12,l}\}_{l=1}^{L_d}$ and $\{\mathcal{V}_{21,l}\}_{l=1}^{L_d}$, respectively, which are described below.

$$\vartheta_{12,l} : \mathcal{Y}_1^n \times \mathcal{V}_{21,1} \times \ldots \times \mathcal{V}_{21,l-1} \to \mathcal{V}_{12,l},$$
$$V_{12,l} = \vartheta_{12,l}(Y_1^n, V_{21}^{l-1}),$$
$$\vartheta_{21,l} : \mathcal{Y}_2^n \times \mathcal{V}_{12,1} \times \ldots \times \mathcal{V}_{12,l-1} \to \mathcal{V}_{21,l},$$
$$V_{21,l} = \vartheta_{21,l}(Y_2^n, V_{12}^{l-1})$$

The conference is said to be $(D_{12}, D_{21})$-permissible if

$$\sum_{l=1}^{L_d} \log\|\mathcal{V}_{12,l}\| \leq nD_{12}, \qquad \sum_{l=1}^{L_d} \log\|\mathcal{V}_{21,l}\| \leq nD_{21}$$
(2)

The receivers exchange information by holding $(D_{12}, D_{21})$-permissible conference. After the conference the first receiver knows the sequence $V_{21}^{L_d} = (V_{21,1}, V_{21,2}, \ldots, V_{21,L_d})$ and the second receiver knows the sequence $V_{12}^{L_d} = (V_{12,1}, \ldots, V_{12,L_d})$.

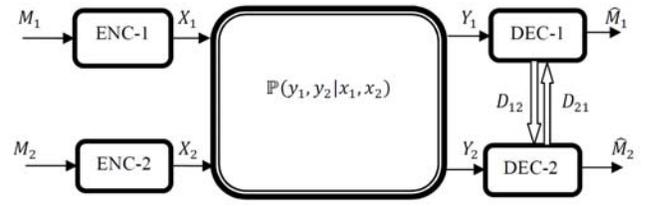

Fig. 1: Interference Channel with Conferencing Decoders.

Lastly, the code includes two decoder functions as follows:

$$\Delta_1 : \mathcal{Y}_1^n \times \mathcal{V}_{21}^{L_d} \to \mathcal{M}_1, \qquad \widehat{M}_1 = \Delta_1(Y_1^n \times V_{21}^{L_d})$$
$$\Delta_2 : \mathcal{Y}_2^n \times \mathcal{V}_{12}^{L_d} \to \mathcal{M}_2, \qquad \widehat{M}_2 = \Delta_2(Y_2^n \times V_{12}^{L_d})$$

Thus, each decoder decodes its message by the respective decoder function.

The capacity region for the two-user IC with conferencing decoders is defined as usual.

## III. MAIN RESULTS

This section is divided into two parts. We present our results for the general two-user IC with cooperative users in Part III.A. Then in Part III.B, we specifically consider the Gaussian channel given in (1).

### A. General IC with Conferencing Decoders

We begin by presenting our novel outer bound for the general two-user IC with conferencing decoders.

Define $\mathfrak{R}_o^{IC \to CD}$ as the union of all rate pairs $(R_1, R_2) \in \mathbb{R}_+^2$ such that

$$R_1 \leq \min\{I(U, X_1; Y_1 | Q) + D_{21}, I(X_1; Y_1 | X_2, Q) + D_{21}\}$$
$$R_1 \leq I(X_1; Y_1 | Y_2, X_2, V, Q) + I(X_1; Y_2 | X_2, Q)$$
$$R_1 \leq I(X_1; Y_2 | Y_1, X_2, V, Q) + I(X_1; Y_1 | X_2, Q)$$
$$R_2 \leq \min\{I(V, X_2; Y_2 | Q) + D_{12}, I(X_2; Y_2 | X_1, Q) + D_{12}\}$$
$$R_2 \leq I(X_2; Y_2 | Y_1, X_1, U, Q) + I(X_2; Y_1 | X_1, Q)$$
$$R_2 \leq I(X_2; Y_1 | Y_2, X_1, U, Q) + I(X_2; Y_2 | X_1, Q)$$
$$R_1 + R_2 \leq I(X_1; Y_1 | V, X_2, Q) + I(V, X_2; Y_2 | Q) + D_{12} + D_{21}$$
$$R_1 + R_2 \leq I(X_2; Y_2 | U, X_1, Q) + I(U, X_1; Y_1 | Q) + D_{12} + D_{21}$$
$$R_1 + R_2 \leq I(X_1; Y_1 | Y_2, X_2, V, Q) + I(X_1, X_2; Y_2 | Q) + D_{12}$$
$$R_1 + R_2 \leq I(X_2; Y_2 | Y_1, X_1, U, Q) + I(X_1, X_2; Y_1 | Q) + D_{21}$$
$$R_1 + R_2 \leq I(X_1, X_2; Y_1, Y_2 | Q)$$
(3)

for some joint PDFs $P_Q P_{X_1 | Q} P_{X_2 | Q} P_{UV | X_1 X_2 Q}$. The following theorem holds.

**Theorem 1.** *The set $\mathfrak{R}_o^{IC \to CD}$ constitutes an outer bound on the capacity region of the two-user IC with decoders connected by the conferencing links of capacities $D_{12}$ and $D_{21}$, as shown in Fig. 1.*

*Proof of Theorem 1:* The proof is given in Appendix I. ∎

Next, using the outer bound (3), we prove four capacity results for IC with unidirectional conferencing between receivers. We highlight that a conferencing link (between receivers) may be utilized to provide one receiver with information about its corresponding signal or its non-corresponding signal (the interference). Our following theorems reveal that both strategies can be helpful to achieve the capacity of the channel.

**Theorem 2.** *For the two-user IC with unidirectional conferencing between decoders, where $D_{21} = 0$, if*

$$I(X_1; Y_1|X_2) \leq I(X_1; Y_2|X_2) \quad \text{for all } P_{X_1}P_{X_2}$$
$$X_2 \to Y_1, X_1 \to Y_2 \quad \text{(Markov chain)}$$
(4)

*then, the outer bound (3) is optimal. The capacity region is given by the union of all $(R_1, R_2) \in \mathbb{R}_+^2$ such that:*

$$R_1 \leq I(X_1; Y_1|X_2, Q),$$
$$R_2 \leq \min\{I(X_2; Y_2|X_1, Q) + D_{12}, I(X_2; Y_1|X_1, Q)\}$$
$$R_1 + R_2 \leq \min\{I(X_1, X_2; Y_2|Q) + D_{12}, I(X_1, X_2; Y_1|Q)\}$$
(5)

*for some joint PDFs $P_Q P_{X_1|Q} P_{X_2|Q}$.*

*Proof of Theorem 2:* Let first prove the achievability of (5). Without loss of generality, assume that the time-sharing variable is null $Q \cong \emptyset$. We present a coding scheme in which both messages are decoded at both receivers. Consider the independent random variables $M_1$ and $M_2$ uniformly distributed over the sets $[1: 2^{nR_1}]$ and $[1: 2^{nR_2}]$, respectively. Partition the set $[1: 2^{nR_2}]$ into $2^{nR_{12}}$ cells each containing $2^{n(R_2 - R_{12})}$ elements, where $R_{12} = \min\{R_2, D_{12}\}$. Now label the cells by $c \in [1: 2^{nR_{12}}]$ and the elements inside each cell by $\kappa \in [1: 2^{n(R_2 - R_{12})}]$. Accordingly, we have $c(M_2) = \theta$ if $M_2$ is inside the cell $\theta$, and $\kappa(M_2) = \beta$ if $M_2$ is the $\beta^{th}$ element of the cell that it belongs to.

Encoding at the transmitters is similar to a MAC. For decoding, the first receiver decodes both messages $M_1$ and $M_2$, by exploring all codewords $X_1^n$ and $X_2^n$ which are jointly typical with its received sequence $Y_1^n$. This receiver then sends the cell index of the estimated message of the second transmitter, i.e. $c(\widehat{M}_2)$, to the receiver $Y_2$ by holding a $(D_{12}, 0)$-permissible conference. The second receiver applies a jointly typical decoder to decode the messages, with the caveat that the cell which $M_2$ belongs to is now known. Clearly, given $c(M_2)$, the second receiver detects the message $M_1$ and $\kappa(M_2)$ by exploring codewords $X_1^n$ and $X_2^n$ which are jointly typical with its received sequence $Y_2^n$. One can easily show that under the conditions (4), this coding scheme yields the achievability of the rate region (5).

Next, using the outer bound (3), we show that under the conditions (4), the achievable rate region (5) is in fact optimal. Based on (3) for $D_{21} = 0$ we have:

$$R_2 \leq I(X_2; Y_2|Y_1, X_1, U, Q) + I(X_2; Y_1|X_1, Q)$$
$$\stackrel{a}{=} I(X_2; Y_1|X_1, Q)$$

$$R_1 + R_2 \leq I(X_1; Y_1|V, X_2, Q) + I(V, X_2; Y_2|Q) + D_{12}$$
$$\stackrel{b}{\leq} I(X_1; Y_2|V, X_2, Q) + I(V, X_2; Y_2|Q) + D_{12}$$
$$= I(X_1, X_2; Y_2|Q) + D_{12}$$

$$R_1 + R_2 \leq I(X_2; Y_2|Y_1, X_1, U, Q) + I(X_1, X_2; Y_1|Q)$$
$$\stackrel{c}{=} I(X_1, X_2; Y_1|Q)$$

where equalities (a) and (c) are due to the second condition of (4) (given $X_1$, $Y_2$ is a degraded version of $Y_1$), and inequality (b) is due to the first condition of (4) (see [21-22]). Note that the other constraints of (5) are directly given by (3) when $D_{21} = 0$. The proof is thus complete. ∎

**Corollary 1.** *Consider the following Gaussian IC with unidirectional conferencing between decoders ($D_{21} = 0$).*

$$Y_1 = S_{11}X_1 + S_{12}X_2 + Z_1$$
$$Y_2 = S_{21}X_1 + S_{22}Y_1 + Z_2$$
(6)

*where $Z_1$ and $Z_2$ are independent unit-variance Gaussian noises. If $\frac{(S_{11}^2 - S_{21}^2)}{2S_{11}S_{21}} \leq S_{22}$, then the capacity region is given by (5).*

*Proof of Corollary 1:* First note that the channel (6) satisfies the second condition of (4) by definition. Moreover, one can easily see that for this channel the first condition of (4) is equivalent to $\frac{(S_{11}^2 - S_{21}^2)}{2S_{11}S_{21}} \leq S_{22}$. Therefore, we can apply the result of Theorem 2. ∎

Based on Theorem 2, for the channel satisfying the conditions (4) the optimal scheme to achieve the capacity region is to decode both messages at both receivers and the optimal cooperation strategy is to provide one receiver with information about its corresponding signal via the conferencing link. In fact, the conditions (4) could be interpreted as a strong interference regime for the IC with unidirectional cooperation between receivers. Note that if the channel satisfies (4), it will also satisfy the standard strong interference regime [16] as well.

**Theorem 3.** *For the two-user IC with unidirectional conferencing between decoders, where $D_{21} = 0$, if*

$$I(V; Y_2|X_2) \leq I(V; Y_1|X_2) \quad \text{for all } P_{X_1}P_{X_2}P_{V|X_1X_2}$$
$$X_2 \to Y_1, X_1 \to Y_2 \quad \text{(Markov chain)}$$
(7)

*then the outer bound (3) is sum-rate optimal and the sum-capacity is given by*

$$\min_{P_Q P_{X_1|Q} P_{X_2|Q}} \left\{ \begin{array}{c} I(X_1; Y_1|X_2, Q) + I(X_2; Y_2|Q) + D_{12}, \\ I(X_1, X_2; Y_1|Q) \end{array} \right\}$$
(8)

*Proof of Theorem 3:* The coding scheme that achieves the sum-rate (8) is similar to that given in the proof of Theorem 2, except for the decoding of the second receiver. Here, the second receiver only decodes its own signal. Given $c(M_2)$, the second receiver detects $\kappa(M_2)$ by exploring codewords $X_2^n$ which are jointly typical with its received sequence $Y_2^n$. One can see that the sum-rate (8) is achieved by this scheme. Now consider the outer bound (3) where $D_{21} = 0$. Under the conditions (7), we have:

$$R_1 + R_2 \leq I(X_1; Y_1|V, X_2, Q) + I(V, X_2; Y_2|Q) + D_{12}$$
$$= I(X_1; Y_1|V, X_2, Q) + I(V; Y_2|X_2, Q) + I(X_2; Y_2|Q) + D_{12}$$
$$\stackrel{a}{\leq} I(X_1; Y_1|V, X_2, Q) + I(V; Y_1|X_2, Q) + I(X_2; Y_2|Q) + D_{12}$$
$$= I(X_1; Y_1|X_2, Q) + I(X_2; Y_2|Q) + D_{12}$$
(9)

where inequality (a) is due to the second condition of (7). Moreover,

$$R_1 + R_2 \leq I(X_2; Y_2|Y_1, X_1, U, Q) + I(X_1, X_2; Y_1|Q)$$
$$\stackrel{b}{=} I(X_1, X_2; Y_1|Q) \quad (10)$$

where equality (b) holds because of the second condition of (7), i.e., given $X_1$, $Y_2$ is a degraded version of $Y_1$, and thereby the first mutual information on the left side of (b) is zero. Therefore, (8) is in fact the sum-rate capacity of the channel and the proof is thus complete. ∎

***Corollary 2.*** Consider the Gaussian IC given in (6) with unidirectional conferencing between decoders, where $D_{21} = 0$. If $\frac{(S_{11}^2 - S_{21}^2)}{2S_{11}S_{21}} \geq S_{22}$, then the sum-rate capacity is given by (8).

*Proof of Corollary 2:* The Gaussian channel (6) satisfies the second condition of (7) by definition. Furthermore, for this channel the first condition of (7) is equivalent to $\frac{(S_{11}^2 - S_{21}^2)}{2S_{11}S_{21}} \geq S_{22}$. Thus, we can apply the result of Theorem 3. ∎

According to Theorem 3, for the channel given in (7) the optimal scheme to achieve the sum-rate capacity is to decode interference at the receiver which is the source of the conferencing link, and to treat interference as noise at the receiver which is the destination of the conferencing link. Moreover, the optimal cooperation strategy is to provide the receiver that treats interference as noise with information about its corresponding signal via the conferencing link. The regime (7) could be indeed interpreted as a mixed interference regime for the IC with unidirectional cooperation between receivers.

In the next theorem, we derive another mixed interference regime for the channel where, unlike Theorem 3, the optimal scheme to achieve the sum-capacity is to treat interference as noise at the receiver which is the source of the conferencing link and to decode interference at the one which is the destination of the conferencing link; also, the optimal cooperation strategy is to provide the receiver that decodes interference with information on its non-corresponding signal (the interference) via the conferencing link.

***Theorem 4.*** For the two-user IC with unidirectional conferencing between receivers, where $D_{21} = 0$, if

$$I(X_1; Y_1|X_2) \leq I(X_1; Y_2|X_2) \quad \text{for all } P_{X_1}P_{X_2}$$
$$X_2 \to Y_2, X_1 \to Y_1 \quad \text{(Markov chain)} \quad (11)$$

then, the outer bound (3) is sum-rate optimal and the sum-rate capacity is given by

$$\min_{P_Q P_{X_1|Q} P_{X_2|Q}} \begin{Bmatrix} I(X_2; Y_2|X_1, Q) + I(X_1; Y_1|Q), \\ I(X_1, X_2; Y_2|Q) + D_{12} \end{Bmatrix} \quad (12)$$

*Proof of Theorem 4:* The achievability of (12) is indeed derived by treating interference as noise at the first receiver and decoding interference at the second receiver. Moreover, the conferencing link is used to provide information about the interference for the second receiver. Let assume $Q \equiv \emptyset$. Consider two independent messages $M_1$ and $M_2$, uniformly distributed over the sets $[1: 2^{nR_1}]$ and $[1: 2^{nR_2}]$, respectively.

Partition the set $[1: 2^{nR_1}]$ into $2^{nR_{12}}$ cells each containing $2^{n(R_1 - R_{12})}$ elements, where $R_{12} = \min\{R_1, D_{12}\}$. Now label the cells by $c \in [1: 2^{n(R_1 - R_{12})}]$ and the elements inside each cell by $\kappa \in [1: 2^{nR_{12}}]$. Accordingly, we have $c(M_1) = \alpha$ if $M_1$ is inside the cell $\alpha$, and $\kappa(M_1) = \beta$ if $M_1$ is $\beta^{th}$ element of the cell that it belongs to.

Encoding at the transmitters is similar to a MAC. For decoding, the first receiver simply decodes its own message by exploring all codewords $X_1^n$ which are jointly typical with its received sequence $Y_1^n$. This receiver then sends the cell index of the estimated message, i.e. $c(\hat{M}_1)$, to the second receiver by holding a $(D_{12}, 0)$-permissible conference. The second receiver applies a jointly typical decoder to decode both messages with the caveat that the cell index which $M_1$ belongs to is known. Clearly, given $c(M_1)$, the second receiver detects $\kappa(M_1)$ and $M_2$ by exploring codewords $X_1^n$ and $X_2^n$ which are jointly typical with its received sequence $Y_2^n$. One can easily show that this scheme yields the achievable sum-rate (12).

Next using our outer bound (3) we prove that under the conditions (11), the sum-rate capacity of the channel is bounded by (12). Based on (3), when $D_{21} = 0$, we have:

$$R_1 + R_2 \leq I(X_2; Y_2|Y_1, X_1, U, Q) + I(X_1, X_2; Y_1|Q)$$
$$= I(X_2; Y_1, Y_2|X_1, U, Q) - I(X_2; Y_1|X_1, U, Q)$$
$$\quad + I(X_2; Y_1|X_1, Q) + I(X_1; Y_1|Q)$$
$$= I(X_2; Y_1, Y_2|X_1, U, Q) + I(U; Y_1|X_1, Q)$$
$$\quad + I(X_1; Y_1|Q)$$
$$= I(X_2; Y_2|X_1, U, Q) + I(X_2; Y_1|Y_2, X_1, U, Q)$$
$$\quad + I(U; Y_1|X_1, Q) + I(X_1; Y_1|Q)$$
$$\stackrel{a}{=} I(X_2; Y_2|X_1, U, Q) + I(U; Y_1|X_1, Q) + I(X_1; Y_1|Q)$$
$$\stackrel{b}{\leq} I(X_2; Y_2|X_1, U, Q) + I(U; Y_2|X_1, Q) + I(X_1; Y_1|Q)$$
$$= I(X_2; Y_2|X_1, Q) + I(X_1; Y_1|Q)$$

where equality (a) holds because by the Markov chain given in (11), the second mutual information on the left side of (a) is zero; similarly, inequality (b) holds because the Markov chain in (11) implies that $I(U; Y_1|X_1, Q) \leq I(U; Y_2|X_1, Q)$. Moreover, we have:

$$R_1 + R_2 \leq I(X_1; Y_1|V, X_2, Q) + I(V, X_2; Y_2|Q) + D_{12}$$
$$\stackrel{c}{\leq} I(X_1; Y_2|V, X_2, Q) + I(V, X_2; Y_2|Q) + D_{12}$$
$$= I(X_1, X_2; Y_2|Q) + D_{12}$$

where inequality (c) is due to the first condition of (11), (see [21-22]). The proof is thus complete. ∎

***Corollary 3.*** Consider the following Gaussian IC with unidirectional conferencing between decoders ($D_{21} = 0$).

$$Y_1 = S_{11}X_1 + S_{12}X_2 + Z_1$$
$$Y_2 = S_{21}X_1 + S_{22}X_2 + Z_2 \quad (13)$$

where $Z_1$ and $Z_2$ are independent unit-variance Gaussian noises. If $\frac{(S_{21}^2 - S_{11}^2)}{2S_{11}S_{21}} \geq S_{12}$, then the sum-rate capacity is given by (12).

*Proof of Corollary 3:* The Gaussian channel (13) satisfies the second condition of (11) by definition. Moreover, for this channel the first condition of (11) is equivalent to $\frac{(S_{21}^2 - S_{11}^2)}{2S_{11}S_{21}} \geq S_{12}$. Thereby, we can apply the result of Theorem 4. ∎

Finally, we characterize the capacity region of the one-sided IC with unidirectional conferencing from the non-interfered receiver to the interfered one in the strong interference regime. This result is given in the following theorem.

**Theorem 5.** *Consider the two-user one-sided IC where $\mathbb{P}(y_1, y_2 | x_1, x_2) = \mathbb{P}(y_1 | x_1) \mathbb{P}(y_2 | x_1, x_2)$. For the channel with unidirection conferencing between receivers, where $Y_1$ is connected to $Y_2$ by a conferencing link of capacity $D_{12}$, if*

$$I(X_1; Y_1 | X_2) \leq I(X_1; Y_2 | X_2) \quad \text{for all } P_{X_1} P_{X_2} \tag{14}$$

*then the outer bound (3) is optimal and the capcity region is given by the union of all rate pairs $(R_1, R_2) \in \mathbb{R}_+^2$ such that*

$$\begin{aligned}
R_1 &\leq I(X_1; Y_1 | Q), \\
R_2 &\leq I(X_2; Y_2 | X_1, Q) \\
R_1 + R_2 &\leq I(X_1, X_2; Y_2 | Q) + D_{12}
\end{aligned} \tag{15}$$

*for some joint PDFs $P_Q P_{X_1 | Q} P_{X_2 | Q}$.*

*Proof of Theorem 5:* The achievability proof is similar to the one presented in Theorem 4. The first receiver simply decodes its own message while the second receiver jointly decodes both messages. The conferencing link is utilized to provide information about the interference (non-desired signal) for the second receiver. For the one-sided channel, this scheme achieves the rate region (15). Then we prove the converse part. Based on (3), when $D_{21} = 0$, we have:

$$\begin{aligned}
R_1 &\leq I(X_1; Y_1 | X_2, Q) = I(X_1; Y_1 | Q) \\
R_2 &\leq I(X_2; Y_1 | Y_2, X_1, U, Q) + I(X_2; Y_2 | X_1, Q) \\
&\stackrel{a}{=} I(X_2; Y_2 | X_1, Q) \\
R_1 + R_2 &\leq I(X_1; Y_1 | V, X_2, Q) + I(V, X_2; Y_2 | Q) + D_{12} \\
&\stackrel{b}{\leq} I(X_1; Y_2 | V, X_2, Q) + I(V, X_2; Y_2 | Q) + D_{12} \\
&= I(X_1, X_2; Y_2 | Q) + D_{12}
\end{aligned}$$

where (a) holds because for the one-sided IC, the first mutual information on the left side of (a) is zero; the inequality (b) is due to the condition (14) (see [21-22]). Thus, the proof is complete. ∎

***Corollary 4.*** *Consider the Gaussian one-sided IC which is given by $S_{12} = 0$ in (1). If $S_{21} \geq S_{11}$, then the capacity region of the channel with unidirectional conferencing between receivers is given by (15). This recovers a result of [14, Th.1].*

Based on the proof of Theorem 5, the optimal coding scheme to achieve the capacity region (15) is to decode both messages at the interfered receiver and the optimal cooperation strategy is to provide the interfered receiver with information about its non-corresponding signal (the interference) via the conferencing link.

### B. Gaussian IC with Conferencing Receivers

In this section, we show that for the Gaussian IC one can tighten the outer bound by utilizing genie-aided techniques as well. Consider the two-user Gaussian IC in (1) with decoders connected by conferencing links of capacities $D_{12}$ and $D_{21}$. Define genie signals $G_1, G_2, \tilde{G}_1$, and $\tilde{G}_2$ as follows:

$$\begin{aligned}
G_1 &\triangleq S_{21} X_1 + Z_2 \\
G_2 &\triangleq S_{12} X_2 + Z_1 \\
\tilde{G}_1 &\triangleq S_{21} X_1 + \tilde{Z}_2 \\
\tilde{G}_2 &\triangleq S_{12} X_2 + \tilde{Z}_1
\end{aligned} \tag{16}$$

where $\tilde{Z}_1$ and $\tilde{Z}_2$ are unit-variance Gaussian noises independent of other variables. Let $\mathfrak{R}_{o,(UV)}^{GIC \to CD}$ denote the set of all rate pairs $(R_1, R_2) \in \mathbb{R}_+^2$ which satisfy the constraints (3) as well as the following:

$$\begin{aligned}
R_1 + R_2 &\leq I(X_1, X_2; Y_1 | G_1, Q) + I(X_1, X_2; Y_2 | G_2, Q) \\
&\quad + D_{12} + D_{21} \\
2R_1 + R_2 &\leq I(X_1; Y_1 | V, X_2, Q) + I(V, X_2; Y_2 | G_2, Q) \\
&\quad + I(X_1, X_2; Y_1 | Q) + D_{12} + 2D_{21} \\
R_1 + 2R_2 &\leq I(X_2; Y_2 | U, X_1, Q) + I(U, X_1; Y_1 | G_1, Q) \\
&\quad + I(X_1, X_2; Y_2 | Q) + 2D_{12} + D_{21} \\
2R_1 + R_2 &\leq I(X_1; Y_1, Y_2 | V, X_2, Q) + I(V, X_2; Y_2 | G_2, Q) \\
&\quad + I(X_1, X_2; Y_1 | Q) + D_{12} + D_{21} \\
R_1 + 2R_2 &\leq I(X_2; Y_1, Y_2 | U, X_1, Q) + I(U, X_1; Y_1 | G_1, Q) \\
&\quad + I(X_1, X_2; Y_2 | Q) + D_{12} + D_{21} \\
2R_1 + R_2 &\leq I(X_1, X_2; Y_1, Y_2 | \tilde{G}_2, Q) + I(X_1, X_2; Y_1 | Q) + D_{21} \\
R_1 + 2R_2 &\leq I(X_1, X_2; Y_1, Y_2 | \tilde{G}_1, Q) + I(X_1, X_2; Y_2 | Q) + D_{12}
\end{aligned} \tag{17}$$

for some joint PDFs $P_Q P_{X_1 | Q} P_{X_2 | Q} P_{UV | X_1 X_2 Q}$. The following theorem holds.

**Theorem 6.** *The set $\mathfrak{R}_{o,(UV)}^{GIC \to CD}$ constitutes an outer bound on the capacity region of the two-user Gaussian IC (1) with conferencing decoders.*

*Proof of Theorem 6.* Refer to Appendix II. ∎

In the following theorem, we present an explicit characterization of the outer bound given in Theorem 6. For this purpose, we indeed apply several novel techniques to optimize the bound over its auxiliary random variables.

**Theorem 7:** *Let $\mathfrak{R}_o^{GIC \to CD}$ denote the set of all rate pairs $(R_1, R_2) \in \mathbb{R}_+^2$ which satisfy the constraints (18) given on the top of the next page for some $\alpha, \beta \in [0,1]$. The set $\mathfrak{R}_o^{GIC \to CD}$ constitutes an outer bound on the capacity region of the Gaussian IC (1) with decoders connected by the conferencing links of capacities $D_{12}$ and $D_{21}$, as shown in Fig. 1.*

*Proof of Theorem 7.* We need to optimize the outer bound established in Theorem 6 over the auxiliary random variables $U$ and $V$, which is indeed a complicated problem. To solve it, we apply novel techniques including several subtle applications of the entropy power inequality. Let present our approach. First note that some of the mutual information functions given in (3) and (17) can be re-written as follows:

$$R_1 \leq \min\left\{\psi\left(\frac{|S_{11}|^2 P_1 + |S_{12}|^2(1-\alpha)P_2}{|S_{12}|^2 \alpha P_2 + 1}\right) + D_{21}, \psi(|S_{11}|^2 P_1) + D_{21}\right\} \tag{1-18}$$

$$R_1 \leq \min\left\{\psi\left(\frac{|S_{11}|^2 \beta P_1}{|S_{21}|^2 \beta P_1 + 1}\right) + \psi(|S_{21}|^2 P_1), \psi\left(\frac{|S_{21}|^2 \beta P_1}{|S_{11}|^2 \beta P_1 + 1}\right) + \psi(|S_{11}|^2 P_1)\right\} \tag{2-18}$$

$$R_2 \leq \min\left\{\psi\left(\frac{|S_{21}|^2(1-\beta)P_1 + |S_{22}|^2 P_2}{|S_{21}|^2 \beta P_1 + 1}\right) + D_{12}, \psi(|S_{22}|^2 P_2) + D_{12}\right\} \tag{3-18}$$

$$R_2 \leq \min\left\{\psi\left(\frac{|S_{22}|^2 \alpha P_2}{|S_{12}|^2 \alpha P_2 + 1}\right) + \psi(|S_{12}|^2 P_2), \psi\left(\frac{|S_{12}|^2 \alpha P_2}{|S_{22}|^2 \alpha P_2 + 1}\right) + \psi(|S_{22}|^2 P_2)\right\} \tag{4-18}$$

$$R_1 + R_2 \leq \left(\psi(|S_{11}|^2 \beta P_1) + \psi\left(\frac{|S_{21}|^2(1-\beta)P_1 + |S_{22}|^2 P_2}{|S_{21}|^2 \beta P_1 + 1}\right)\right)\mathbb{1}(|S_{21}| < |S_{11}|)$$
$$+ \psi(|S_{21}|^2 P_1 + |S_{22}|^2 P_2)\mathbb{1}(|S_{21}| \geq |S_{11}|) + D_{12} + D_{21} \tag{5-18}$$

$$R_1 + R_2 \leq \left(\psi(|S_{22}|^2 \alpha P_2) + \psi\left(\frac{|S_{11}|^2 P_1 + |S_{12}|^2(1-\alpha)P_2}{|S_{12}|^2 \alpha P_2 + 1}\right)\right)\mathbb{1}(|S_{12}| < |S_{22}|)$$
$$+ \psi(|S_{11}|^2 P_1 + |S_{12}|^2 P_2)\mathbb{1}(|S_{12}| \geq |S_{22}|) + D_{12} + D_{21} \tag{6-18}$$

$$R_1 + R_2 \leq \psi\left(\frac{|S_{11}|^2 \beta P_1}{|S_{21}|^2 \beta P_1 + 1}\right) + \psi(|S_{21}|^2 P_1 + |S_{22}|^2 P_2) + D_{12} \tag{7-18}$$

$$R_1 + R_2 \leq \psi\left(\frac{|S_{22}|^2 \alpha P_2}{|S_{12}|^2 \alpha P_2 + 1}\right) + \psi(|S_{11}|^2 P_1 + |S_{12}|^2 P_2) + D_{21} \tag{8-18}$$

$$R_1 + R_2 \leq \psi(|S_{11}|^2 P_1 + |S_{12}|^2 P_2 + |S_{21}|^2 P_1 + |S_{22}|^2 P_2 + |S_{11}S_{22} - S_{12}S_{21}|^2 P_1 P_2) \tag{9-18}$$

$$R_1 + R_2 \leq \psi\left(|S_{12}|^2 P_2 + \frac{|S_{11}|^2 P_1}{|S_{21}|^2 P_1 + 1}\right) + \psi\left(|S_{21}|^2 P_1 + \frac{|S_{22}|^2 P_2}{|S_{12}|^2 P_2 + 1}\right) + D_{12} + D_{21} \tag{10-18}$$

$$2R_1 + R_2 \leq \left(\psi(|S_{11}|^2 \beta P_1) - \psi(|S_{21}|^2 \beta P_1)\right)\mathbb{1}(|S_{21}| < |S_{11}|) + \psi\left(|S_{21}|^2 P_1 + \frac{|S_{22}|^2 P_2}{|S_{12}|^2 P_2 + 1}\right)$$
$$+ \psi(|S_{11}|^2 P_1 + |S_{12}|^2 P_2) + D_{12} + 2D_{21} \tag{11-18}$$

$$R_1 + 2R_2 \leq \left(\psi(|S_{22}|^2 \alpha P_2) - \psi(|S_{12}|^2 \alpha P_2)\right)\mathbb{1}(|S_{12}| < |S_{22}|) + \psi\left(|S_{12}|^2 P_2 + \frac{|S_{11}|^2 P_1}{|S_{21}|^2 P_1 + 1}\right)$$
$$+ \psi(|S_{21}|^2 P_1 + |S_{22}|^2 P_2) + 2D_{12} + D_{21} \tag{12-18}$$

$$2R_1 + R_2 \leq \psi((|S_{11}|^2 + |S_{21}|^2)\beta P_1) + \psi\left(\frac{|S_{21}|^2(1-\beta)P_1}{1 + |S_{21}|^2 \beta P_1} + \frac{|S_{22}|^2 P_2}{(|S_{12}|^2 P_2 + 1)(1 + |S_{21}|^2 \beta P_1)}\right)$$
$$+ \psi(|S_{11}|^2 P_1 + |S_{12}|^2 P_2) + D_{12} + D_{21} \tag{13-18}$$

$$R_1 + 2R_2 \leq \psi((|S_{12}|^2 + |S_{22}|^2)\alpha P_2) + \psi\left(\frac{|S_{12}|^2(1-\alpha)P_2}{1 + |S_{12}|^2 \alpha P_2} + \frac{|S_{11}|^2 P_1}{(|S_{21}|^2 P_1 + 1)(1 + |S_{12}|^2 \alpha P_2)}\right)$$
$$+ \psi(|S_{21}|^2 P_1 + |S_{22}|^2 P_2) + D_{12} + D_{21} \tag{14-18}$$

$$2R_1 + R_2 \leq \psi\left(|S_{11}|^2 P_1 + \frac{|S_{12}|^2 P_2}{1 + |S_{12}|^2 P_2} + |S_{21}|^2 P_1 + \frac{|S_{22}|^2 P_2}{1 + |S_{12}|^2 P_2} + \frac{|S_{11}S_{22} - S_{12}S_{21}|^2 P_1 P_2}{1 + |S_{12}|^2 P_2}\right)$$
$$+ \psi(|S_{11}|^2 P_1 + |S_{12}|^2 P_2) + D_{21} \tag{15-18}$$

$$R_1 + 2R_2 \leq \psi\left(\frac{|S_{11}|^2 P_1}{1 + |S_{21}|^2 P_1} + |S_{12}|^2 P_2 + \frac{|S_{21}|^2 P_1}{1 + |S_{21}|^2 P_1} + |S_{22}|^2 P_2 + \frac{|S_{11}S_{22} - S_{12}S_{21}|^2 P_1 P_2}{1 + |S_{21}|^2 P_1}\right)$$
$$+ \psi(|S_{21}|^2 P_1 + |S_{22}|^2 P_2) + D_{12} \tag{16-18}$$

$$\tag{18}$$

$I(X_2; Y_2|Y_1, X_1, U, Q)$
$\quad = I(X_2; Y_1, Y_2|U, X_1, Q) - I(X_2; Y_1|U, X_1, Q)$

$I(X_2; Y_1|Y_2, X_1, U, Q)$
$\quad = I(X_2; Y_1, Y_2|U, X_1, Q) - I(X_2; Y_2|U, X_1, Q)$

$I(X_1; Y_1|Y_2, X_2, V, Q)$
$\quad = I(X_1; Y_1, Y_2|V, X_2, Q) - I(X_1; Y_2|V, X_2, Q)$

$I(X_1; Y_2|Y_1, X_2, V, Q)$
$\quad = I(X_1; Y_1, Y_2|V, X_2, Q) - I(X_1; Y_1|V, X_2, Q)$

$$\tag{19}$$

In general it is difficult to directly treat expressions such as $I(X_1; Y_1, Y_2|V, X_2, Q)$ or $I(X_2; Y_1, Y_2|U, X_1, Q)$. To make the problem tractable, we apply the following technique. Let define two new outputs $\hat{Y}_1$ and $\hat{Y}_2$ as follows:

$$\hat{Y}_1 \triangleq \frac{S_{12}Y_1 + S_{22}Y_2}{|S_{12}|^2 + |S_{22}|^2} = \frac{S_{11}S_{12} + S_{21}S_{22}}{|S_{12}|^2 + |S_{22}|^2}X_1 + X_2 + \hat{Z}_1$$

$$\hat{Y}_2 \triangleq \frac{S_{11}Y_1 + S_{21}Y_2}{|S_{11}|^2 + |S_{21}|^2} = X_1 + \frac{S_{11}S_{12} + S_{21}S_{22}}{|S_{11}|^2 + |S_{21}|^2}X_2 + \hat{Z}_2$$

(20)

where $\hat{Z}_1$ and $\hat{Z}_2$ are given as:

$$\hat{Z}_1 \triangleq \frac{S_{12}Z_1 + S_{22}Z_2}{|S_{12}|^2 + |S_{22}|^2}$$

$$\hat{Z}_2 \triangleq \frac{S_{11}Z_1 + S_{21}Z_2}{|S_{11}|^2 + |S_{21}|^2}$$

(21)

It is clear that the mapping from $(Y_1, Y_2)$ to $(\hat{Y}_1, Y_2)$ and also to $(Y_1, \hat{Y}_2)$ is one-to-one. Now we have:

$$Y_1 = S_{11}\hat{Y}_2 + \frac{S_{21}(S_{12}S_{21} - S_{11}S_{22})}{|S_{11}|^2 + |S_{21}|^2}X_2 + \bar{\bar{Z}}_1$$

$$Y_2 = S_{22}\hat{Y}_1 + \frac{S_{12}(S_{12}S_{21} - S_{11}S_{22})}{|S_{12}|^2 + |S_{22}|^2}X_1 + \bar{\bar{Z}}_2$$

(22)

where

$$\bar{\bar{Z}}_1 \triangleq \frac{S_{21}(S_{21}Z_1 - S_{11}Z_2)}{|S_{11}|^2 + |S_{21}|^2}$$

$$\bar{\bar{Z}}_2 \triangleq \frac{S_{12}(S_{12}Z_2 - S_{22}Z_1)}{|S_{12}|^2 + |S_{22}|^2}$$

(23)

One can easily check that $\bar{\bar{Z}}_1$ is independent of $\hat{Z}_1$ and also $\bar{\bar{Z}}_2$ is independent of $\hat{Z}_2$. Therefore, the following equalities hold:

$$I(X_2; Y_1, Y_2|U, X_1, Q) = I(X_2; \hat{Y}_1, Y_2|U, X_1, Q)$$
$$= I(X_2; \hat{Y}_1|U, X_1, Q)$$

$$I(X_1; Y_1, Y_2|V, X_2, Q) = I(X_1; Y_1, \hat{Y}_2|V, X_2, Q)$$
$$= I(X_1; \hat{Y}_2|V, X_2, Q)$$

(24)

for any arbitrary input distributions. Next fix a distribution $P_Q P_{X_1|Q} P_{X_2|Q} P_{UV|X_1X_2Q}$ with $\mathbb{E}[X_i^2] \leq P_i, i = 1,2$. In what follows, we present the optimization for the auxiliary random variable $U$. The optimization over $V$ is derived symmetrically, and therefore we do not present the details to avoid repetition.

Let divide the problem into two different cases. First consider the channel with weak interference at the first receiver where

$$|S_{12}| < |S_{22}|$$

(25)

It is clear that:

$$\frac{1}{2}\log 2\pi e = H(Y_2|U, X_1, X_2, Q)$$
$$\leq H(Y_2|U, X_1, Q)$$
$$= H(S_{22}X_2 + Z_2|U, X_1, Q)$$
$$\leq H(S_{22}X_2 + Z_2|Q) \leq \frac{1}{2}\log 2\pi e(|S_{22}|^2 P_2 + 1)$$

(26)

Comparing the two sides of (26), one can deduce that there is $\alpha \in [0,1]$ such that:

$$H(Y_2|U, X_1, Q) = H(S_{22}X_2 + Z_2|U, X_1, Q)$$
$$= \frac{1}{2}\log 2\pi e(|S_{22}|^2 \alpha P_2 + 1)$$

(27)

Then by considering (27) and also (19) and (24), one can easily verify that the optimization is equivalent to maximize $H(\hat{Y}_1|U, X_1, Q)$ and minimize $H(Y_1|U, X_1, Q)$, simultaneously. For this purpose, we apply the entropy power inequality. This inequality implies that for any arbitrary random variables $X, Z$, and $W$, where $X$ and $Z$ are independent conditioned on $W$, the following holds:

$$\exp(2H(X + Z|W)) \geq \exp(2H(X|W)) + \exp(2H(Z|W))$$

(28)

Therefore, assuming that $H(Z|W)$ is fixed, given $H(X + Z|W)$, one can derive an upper bound on $H(X|W)$, and given $H(X|W)$, one can derive a lower bound on $H(X + Z|W)$. This fact is the essence of our arguments in what follows.

Let $\hat{Z}_1^*$ be a Gaussian random variable, independent of all other variables, with zero mean and a variance equal to $\frac{|S_{12}|^2}{|S_{12}|^2 + |S_{22}|^2}$. One can write:

$$H(\hat{Y}_1|U, X_1, Q)$$
$$= H\left(\frac{S_{12}Y_1 + S_{22}Y_2}{|S_{12}|^2 + |S_{22}|^2}\bigg|U, X_1, Q\right)$$
$$= H(X_2 + \hat{Z}_1|U, X_1, Q)$$
$$= H(S_{22}X_2 + S_{22}\hat{Z}_1|U, X_1, Q) - \frac{1}{2}\log|S_{22}|^2$$
$$\overset{a}{\leq} \frac{1}{2}\log\left(\begin{array}{c}\exp\left(2H(S_{22}X_2 + S_{22}\hat{Z}_1 + \hat{Z}_1^*|U, X_1, Q)\right) \\ -\exp\left(2H(\hat{Z}_1^*|U, X_1, Q)\right)\end{array}\right)$$
$$\quad - \frac{1}{2}\log|S_{22}|^2$$
$$\overset{b}{=} \frac{1}{2}\log\left(\begin{array}{c}\exp(2H(S_{22}X_2 + Z_2|U, X_1, Q)) \\ -\exp\left(2H(\hat{Z}_1^*)\right)\end{array}\right) - \frac{1}{2}\log|S_{22}|^2$$
$$\overset{c}{=} \frac{1}{2}\log\left(\begin{array}{c}\exp(\log 2\pi e(|S_{22}|^2\alpha P_2 + 1)) \\ -\exp\left(\log 2\pi e\left(\frac{|S_{12}|^2}{|S_{12}|^2 + |S_{22}|^2}\right)\right)\end{array}\right)$$
$$\quad - \frac{1}{2}\log|S_{22}|^2$$
$$= \frac{1}{2}\log\left(2\pi e\left(\alpha P_2 + \frac{1}{|S_{12}|^2 + |S_{22}|^2}\right)\right)$$

(29)

where (a) is due to the EPI, (b) holds because $S_{22}\hat{Z}_1 + \hat{Z}_1^*$ is a Gaussian random variable with zero mean and unit variance (the same as $Z_2$), and (c) is given by (27). Next let $Z_2^*$ be a Gaussian random variable, independent of all other variables, with zero mean and a variance equal to $\frac{|S_{22}|^2}{|S_{12}|^2} - 1$. We have:

$$H(Y_1|U, X_1, Q) = H(S_{12}X_2 + Z_1|U, X_1, Q)$$
$$= H\left(S_{22}X_2 + \frac{S_{22}}{S_{12}}Z_1\bigg|U, X_1, Q\right) - \frac{1}{2}\log\frac{|S_{22}|^2}{|S_{12}|^2}$$

$$\stackrel{a}{=} H(S_{22}X_2 + Z_2 + Z_2^* | U, X_1, Q) - \frac{1}{2}\log\frac{|S_{22}|^2}{|S_{12}|^2}$$

$$\stackrel{b}{\geq} \frac{1}{2}\log\begin{pmatrix} \exp(2H(S_{22}X_2 + Z_2|U,X_1,Q)) \\ + \exp(2H(Z_2^*|U,X_1,Q)) \end{pmatrix}$$
$$- \frac{1}{2}\log\frac{|S_{22}|^2}{|S_{12}|^2}$$

$$\stackrel{c}{=} \frac{1}{2}\log\begin{pmatrix} \exp(\log 2\pi e(|S_{22}|^2 \alpha P_2 + 1)) \\ + \exp\left(\log 2\pi e\left(\frac{|S_{22}|^2}{|S_{12}|^2} - 1\right)\right) \end{pmatrix}$$
$$- \frac{1}{2}\log\frac{|S_{22}|^2}{|S_{12}|^2}$$

$$= \frac{1}{2}\log(2\pi e(|S_{12}|^2 \alpha P_2 + 1))$$
(30)

where (a) holds because $Z_2 + Z_2^*$ is a Gaussian random variable with zero mean and a variance equal to $\frac{|S_{22}|^2}{|S_{12}|^2}$, i.e., the same as $\frac{S_{22}}{S_{12}}Z_1$, (b) is due to the EPI, and (c) is given by (27). Thus, we applied the entropy power inequality twice: once to establish an upper bound on $H(\hat{Y}_1|U,X_1,Q)$ as in (29) and once to establish a lower bound on $H(Y_1|U,X_1,Q)$ as in (30). It is important to note that one may also apply the principle of "Gaussian maximizes differential entropy" to obtain an upper bound on $H(\hat{Y}_1|U,X_1,Q)$, however, the upper bound derived by that approach does not necessarily relate to $\alpha$ (which is specifically determined by (27)) and thereby we cannot establish a bound consistent to other entropy functions including the auxiliary random variable $U$.

Let again consider the derivations in (30). One can easily see that all of the relations given in (30) hold only for the channel with weak interference at the first receiver where $|S_{12}| < |S_{22}|$. As the second case, we need to consider the Gaussian IC with strong interference at the first receiver where

$$|S_{12}| \geq |S_{22}|$$
(31)

For this case, the derivations in (30) are no longer valid. The fact is that when $|S_{21}| \geq |S_{11}|$, by fixing $H(Y_2|U,X_1,Q)$ as in (27), we cannot establish a lower bound on $H(Y_1|U,X_1,Q)$ using the EPI because given $X_2$, the output $Y_1$ is not a stochastically degraded version of $Y_2$ anymore. Therefore, we need to change our strategy for the optimization. For this purpose, first note that the strong interference condition (31) implies the following inequality (see [21]):

$$I(X_2;Y_2|U,X_1,Q) \leq I(X_2;Y_1|U,X_1,Q) \text{ for all PDFs } P_{QUX_1X_2}$$
(32)

Considering (32), we can derive:

$$I(X_2;Y_2|U,X_1,Q) + I(U,X_1;Y_1|Q)$$
$$\leq I(X_2;Y_1|U,X_1,Q) + I(U,X_1;Y_1|Q)$$
$$= I(X_1,X_2;Y_1|Q)$$
(33)

$$I(X_2;Y_2|U,X_1,Q) + I(U,X_1;Y_1|G_1,Q)$$
$$\leq I(X_2;Y_1|U,X_1,Q) + I(U,X_1;Y_1|G_1,Q)$$
$$= I(X_1,X_2;Y_1|G_1,Q)$$
(34)

Then, we evaluate $H(\hat{Y}_1|U,X_1,Q)$. We can write:

$$\frac{1}{2}\log\left(2\pi e\left(\frac{1}{|S_{12}|^2 + |S_{22}|^2}\right)\right)$$
$$= H(\hat{Y}_1|U,X_1,X_2,Q)$$
$$\leq H(\hat{Y}_1|U,X_1,Q)$$
$$= H\left(\frac{S_{12}Y_1 + S_{22}Y_2}{|S_{12}|^2 + |S_{22}|^2}\bigg|U,X_1,Q\right)$$
$$= H(X_2 + \hat{Z}_1|U,X_1,Q)$$
$$\leq H(X_2 + \hat{Z}_1|Q)$$
$$\leq \frac{1}{2}\log\left(2\pi e\left(P_2 + \frac{1}{|S_{12}|^2 + |S_{22}|^2}\right)\right)$$
(35)

Comparing the two sides of (35), we can argue that there is $\alpha \in [0,1]$ such that:

$$H(\hat{Y}_1|U,X_1,Q) = \frac{1}{2}\log\left(2\pi e\left(\alpha P_2 + \frac{1}{|S_{12}|^2 + |S_{22}|^2}\right)\right)$$
(36)

Now by substituting (33-34) in (3) and (17) and considering (19) and (24), one can readily verify that the optimization is reduced to minimize $H(Y_1|U,X_1,Q)$ and $H(Y_2|U,X_1,Q)$, simultaneously. Moreover, given $X_1$, both $Y_1$ and $Y_2$ are stochastically degraded versions of $\hat{Y}_1$. Therefore, considering (36), one can successfully apply the EPI to establish lower bounds on $H(Y_1|U,X_1,Q)$ and $H(Y_2|U,X_1,Q)$. Clearly, let $\hat{Z}_1^\nabla$ and $\hat{Z}_1^\Delta$ be two Gaussian random variables, independent of all other variables, with zero mean and variances $\frac{|S_{22}|^2}{|S_{12}|^2(|S_{12}|^2 + |S_{22}|^2)}$ and we have $\frac{|S_{12}|^2}{|S_{22}|^2(|S_{12}|^2 + |S_{22}|^2)}$, respectively. We have:

$$H(Y_1|U,X_1,Q)$$
$$= H(S_{12}X_2 + Z_1|U,X_1,Q)$$
$$= H\left(X_2 + \frac{1}{S_{12}}Z_1\bigg|U,X_1,Q\right) - \frac{1}{2}\log\frac{1}{|S_{12}|^2}$$
$$\stackrel{a}{=} H(X_2 + \hat{Z}_1 + \hat{Z}_1^\nabla|U,X_1,Q) - \frac{1}{2}\log\frac{1}{|S_{12}|^2}$$
$$\stackrel{b}{\geq} \frac{1}{2}\log\begin{pmatrix} \exp\left(2H(X_2 + \hat{Z}_1|U,X_1,Q)\right) \\ + \exp\left(2H(\hat{Z}_1^\nabla|U,X_1,Q)\right) \end{pmatrix} - \frac{1}{2}\log\frac{1}{|S_{12}|^2}$$
$$\stackrel{c}{=} \frac{1}{2}\log\begin{pmatrix} \exp\left(\log 2\pi e\left(\alpha P_2 + \frac{1}{|S_{12}|^2 + |S_{22}|^2}\right)\right) \\ + \exp\left(\log 2\pi e\left(\frac{|S_{22}|^2}{|S_{12}|^2(|S_{12}|^2 + |S_{22}|^2)}\right)\right) \end{pmatrix}$$
$$- \frac{1}{2}\log\frac{1}{|S_{12}|^2}$$
$$= \frac{1}{2}\log(2\pi e(|S_{12}|^2 \alpha P_2 + 1))$$
(37)

where (a) holds because $\hat{Z}_1 + \hat{Z}_1^\nabla$ is a Gaussian variable with zero mean and variance $\frac{1}{|S_{12}|^2}$, i.e., the same as $\frac{1}{S_{12}}Z_1$, (b) is due to the EPI, and (c) is given by (36). Similarly, we can derive:

$$\begin{aligned}
&H(Y_2|U,X_1,Q)\\
&= H(S_{22}X_2 + Z_2|U,X_1,Q)\\
&= H\left(X_2 + \frac{1}{S_{22}}Z_2\Big|U,X_1,Q\right) - \frac{1}{2}\log\frac{1}{|S_{22}|^2}\\
&\stackrel{a}{=} H(X_2 + \hat{Z}_1 + \hat{Z}_1^\Delta|U,X_1,Q) - \frac{1}{2}\log\frac{1}{|S_{22}|^2}\\
&\stackrel{b}{\geq} \frac{1}{2}\log\left(\begin{array}{l}\exp\left(2H(X_2+\hat{Z}_1|U,X_1,Q)\right)\\+\exp\left(2H(\hat{Z}_1^\Delta|U,X_1,Q)\right)\end{array}\right) - \frac{1}{2}\log\frac{1}{|S_{22}|^2}\\
&\stackrel{c}{=} \frac{1}{2}\log\left(\begin{array}{l}\exp\left(\log 2\pi e\left(\alpha P_2 + \frac{1}{|S_{12}|^2+|S_{22}|^2}\right)\right)\\+\exp\left(\log 2\pi e\left(\frac{|S_{12}|^2}{|S_{22}|^2(|S_{12}|^2+|S_{22}|^2)}\right)\right)\end{array}\right)\\
&\quad - \frac{1}{2}\log\frac{1}{|S_{22}|^2}\\
&= \frac{1}{2}\log(2\pi e(|S_{22}|^2\alpha P_2 + 1))
\end{aligned}$$
(38)

where (a) holds because $\hat{Z}_1 + \hat{Z}_1^\Delta$ is a Gaussian variable with zero mean and variance $\frac{1}{|S_{22}|^2}$, i.e., the same as $\frac{1}{S_{22}}Z_2$, (b) is due to the EPI, and (c) is given by (36). Therefore, $H(Y_1|U,X_1,Q)$ and $H(Y_2|U,X_1,Q)$ are minimized by the right side of (37) and (38), respectively. The proof is thus complete. ∎

As indicated earlier, an outer bound was also established by Wang and Tse in [9] for the Gaussian IC with conferencing decoders. In the following, we argue that ours is strictly tighter than that of [9].

***Remark.*** *For all channel parameters, the outer bound $\mathfrak{R}_o^{GIC \to CD}$ given by (18) is strictly tighter than that of [9, Lemma 5.1].*

In fact by a straightforward comparison via simple algebraic computations, one can verify that each of the constraints given in (18) is tighter than the corresponding one of [9, Page 2920].

We conclude our paper by providing some numerical results. In Fig. 1, 2, and 3, we compare our new outer bound for the Gaussian IC with conferencing decoders and that of [9] in the weak, strong, and mixed interference regimes, respectively. As shown in these figures, for all cases our new outer bound could be strictly tighter.

## CONCLUSION

In this paper, we investigated capacity bounds for the two-user Interference Channel (IC) with cooperative receivers via conferencing links of finite capacities. By applying new techniques, we presented novel capacity outer bounds for this channel. Using the outer bounds, we proved several new capacity results are proved for interesting channels with unidirectional cooperation in strong and mixed interference regimes. A fact is that a conferencing link (between receivers) may be utilized to provide one receiver with information about its corresponding signal or its non-corresponding signal (interference). An interesting conclusion of the paper was to show that both of these strategies can be helpful to achieve the capacity of the channel. Finally, for the case of Gaussian IC, we showed that our outer bound is strictly tighter than the previous one derived by Wang and Tse [9].

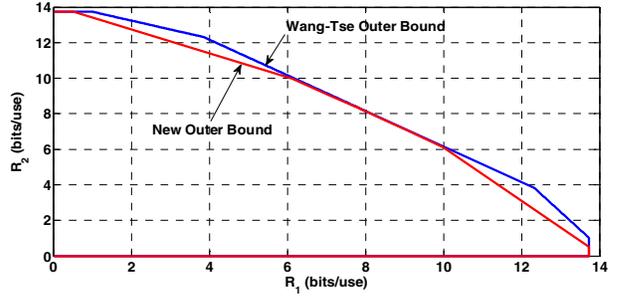

Fig. 2: Comparison of the new outer bound for the Gaussian IC (1) with conferencing decoders and that of [9] in the weak interference regime ($P_1 = P_2 = 1$, $D_{12} = D_{21} = .5$, $S_{11} = S_{22} = 100$, $S_{12} = S_{21} = 60$).

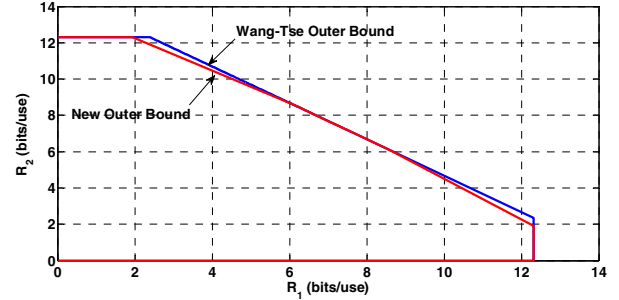

Fig. 3: Comparison of the new outer bound for the Gaussian IC (1) with conferencing decoders and that of [9] in the strong interference regime ($P_1 = P_2 = 1$, $D_{12} = D_{21} = .5$, $S_{11} = S_{22} = 60$, $S_{12} = S_{21} = 100$).

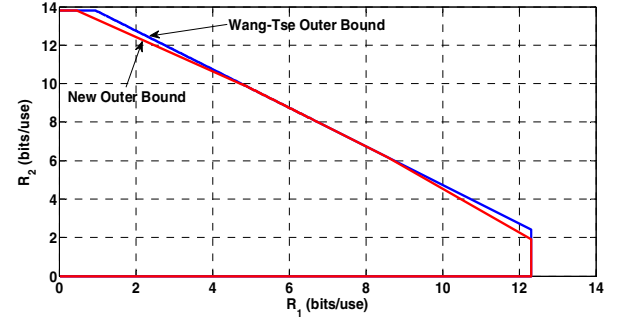

Fig. 4: Comparison of the new outer bound for the Gaussian IC (1) with conferencing decoders and that of [9] in the mixed interference regime ($P_1 = P_2 = 1$, $D_{12} = D_{21} = .5$, $S_{11} = S_{21} = 60$, $S_{12} = S_{22} = 100$).

In our ongoing work, we are investigating the interesting question that whether our new outer bound for the Gaussian channel could be used to obtain a better approximation of the capacity region compared to the result of [9].

## APPENDIX I

### PROOF OF THEOREM 1

Consider a length-$n$ code with vanishing average probability of error. Define new auxiliary random variables:

$$\begin{aligned}
U_t &\triangleq (M_1, Y_2^{t-1}, Y_{1,t+1}^n)\\
V_t &\triangleq (M_2, Y_1^{t-1}, Y_{2,t+1}^n), \quad \text{for} \quad t = 1, \ldots, n
\end{aligned}$$
(A.1)

Let first derive some bounds on $R_1$. By Fano's inequality,

$$nR_1 \leq I(M_1; Y_1^n, V_{21}^{Ld}) + n\epsilon_n^1$$

$$= I(M_1; Y_1^n) + I(M_1; V_{21}^{L_d}|Y_1^n) + n\epsilon_n^1$$

$$\overset{a}{\leq} I(M_1; Y_1^n|M_2) + H(V_{21}^{L_d}) + n\epsilon_n^1$$

$$\overset{b}{\leq} I(M_1; Y_1^n|M_2) + nD_{21} + n\epsilon_n^1$$

$$\leq \sum_{t=1}^{n} I(X_{1,t}; Y_{1,t}|X_{2,t}) + nD_{21} + n\epsilon_n^1$$

(A.2)

where inequality (a) holds because conditioning does not reduce the entropy, and inequality (b) is due to (2). Moreover,

$$nR_1 \leq I(M_1; Y_1^n, V_{21}^{L_d}) + n\epsilon_n^1$$

$$= I(M_1; Y_1^n) + I(M_1; V_{21}^{L_d}|Y_1^n) + n\epsilon_n^1$$

$$\leq \sum_{t=1}^{n} I(M_1; Y_{1,t}|Y_{1,t+1}^n) + H(V_{21}^{L_d}) + n\epsilon_n^1$$

$$\overset{a}{\leq} \sum_{t=1}^{n} I(M_1, Y_{1,t+1}^n, Y_2^{t-1}, X_{1,t}; Y_{1,t}) + H(V_{21}^{L_d}) + n\epsilon_n^1$$

$$\leq \sum_{t=1}^{n} I(U_t, X_{1,t}; Y_{1,t}) + nD_{21} + n\epsilon_n^1$$

(A.3)

where inequality (a) holds because conditioning does not reduce the entropy. Next we derive some bounds on $R_2$. By Fano's inequality we have:

$$nR_2 \leq I(M_2; Y_2^n, V_{12}^{L_d}) + n\epsilon_n^2$$

$$\leq I(M_2; Y_1^n, Y_2^n, V_{12}^{L_d}) + n\epsilon_n^2$$

$$\overset{a}{=} I(M_2; Y_1^n, Y_2^n) + n\epsilon_n^2$$

$$\leq I(M_2; Y_1^n, Y_2^n|M_1) + n\epsilon_n^2$$

(A.4)

where (a) holds because $V_{12}^{L_d}$ is given by a deterministic function of $(Y_1^n, Y_2^n)$. Now consider the mutual information function on the right side of the last inequality of (A.4). We can write:

$$I(M_2; Y_1^n, Y_2^n|M_1) = I(M_2; Y_2^n|M_1) + I(M_2; Y_1^n|Y_2^n, M_1)$$

$$\overset{a}{=} \sum_{t=1}^{n} I(X_{2,t}; Y_{2,t}|X_{1,t}, M_1)$$

$$+ \sum_{t=1}^{n} I(X_{2,t}; Y_{1,t}|Y_{2,t}, X_{1,t}, M_1, Y_2^{t-1}, Y_{2,t+1}^n, Y_{1,t+1}^n)$$

$$= \sum_{t=1}^{n} I(X_{2,t}; Y_{2,t}|X_{1,t}, M_1)$$

$$+ \sum_{t=1}^{n} I(X_{2,t}; Y_{1,t}|Y_{2,t}, X_{1,t}, U_t, Y_{2,t+1}^n)$$

$$\overset{b}{\leq} \sum_{t=1}^{n} I(X_{2,t}; Y_{2,t}|X_{1,t}) + \sum_{t=1}^{n} I(X_{2,t}; Y_{1,t}|Y_{2,t}, X_{1,t}, U_t)$$

(A.5)

where (a) holds because $X_{i,t}$ is given by a deterministic function of $M_i$; the inequality (b) holds because conditioning does not reduce the entropy and also given the inputs $X_{1,t}, X_{2,t}$, the outputs $Y_{1,t}, Y_{2,t}$ are independent of other variables. Similarly, we can derive:

$$I(M_2; Y_1^n, Y_2^n|M_1) = I(M_2; Y_1^n|M_1) + I(M_2; Y_2^n|Y_1^n, M_1)$$

$$\leq \sum_{t=1}^{n} I(X_{2,t}; Y_{1,t}|X_{1,t}) + \sum_{t=1}^{n} I(X_{2,t}; Y_{2,t}|Y_{1,t}, X_{1,t}, U_t)$$

(A.6)

Now by substituting (A.5) and (A.6) in (A.4), we obtain:

$$nR_2 \leq$$

$$\min \begin{cases} \sum_{t=1}^{n} I(X_{2,t}; Y_{2,t}|X_{1,t}) + I(X_{2,t}; Y_{1,t}|Y_{2,t}, X_{1,t}, U_t), \\ \sum_{t=1}^{n} I(X_{2,t}; Y_{1,t}|X_{1,t}) + I(X_{2,t}; Y_{2,t}|Y_{1,t}, X_{1,t}, U_t) \end{cases}$$

(A.7)

Finally, we establish constraints on the sum-rate. Based on Fano's inequality, one can write:

$$n(R_1 + R_2) \leq I(M_1; Y_1^n, V_{21}^{L_d}) + I(M_2; Y_2^n, V_{12}^{L_d}) + n(\epsilon_n^1 + \epsilon_n^2)$$

$$\leq I(M_1; Y_1^n) + I(M_2; Y_2^n) + I(M_1; V_{21}^{L_d}|Y_1^n)$$

$$+ I(M_2; V_{12}^{L_d}|Y_2^n) + n(\epsilon_n^1 + \epsilon_n^2)$$

$$\leq I(M_1; Y_1^n) + I(M_2; Y_2^n|M_1) + H(V_{21}^{L_d})$$

$$+ H(V_{12}^{L_d}) + n(\epsilon_n^1 + \epsilon_n^2)$$

$$\leq I(M_1; Y_1^n) + I(M_2; Y_2^n|M_1)$$

$$+ n(D_{12} + D_{21}) + n(\epsilon_n^1 + \epsilon_n^2)$$

(A.8)

The sum of the two mutual information functions on the right side of (A.8) can be bounded as follows:

$$I(M_1; Y_1^n) + I(M_2; Y_2^n|M_1)$$

$$= \sum_{t=1}^{n} I(M_1; Y_{1,t}|Y_{1,t+1}^n) + \sum_{t=1}^{n} I(M_2; Y_{2,t}|M_1, Y_2^{t-1})$$

$$\leq \sum_{t=1}^{n} I(Y_2^{t-1}, M_1; Y_{1,t}|Y_{1,t+1}^n) - \sum_{t=1}^{n} I(Y_2^{t-1}; Y_{1,t}|M_1, Y_{1,t+1}^n)$$

$$+ \sum_{t=1}^{n} I(Y_{1,t+1}^n, M_2; Y_{2,t}|M_1, Y_2^{t-1})$$

$$= \sum_{t=1}^{n} I(Y_2^{t-1}, M_1; Y_{1,t}|Y_{1,t+1}^n)$$

$$- \sum_{t=1}^{n} I(Y_2^{t-1}; Y_{1,t}|M_1, Y_{1,t+1}^n)$$

$$+ \sum_{t=1}^{n} I(Y_{1,t+1}^n; Y_{2,t}|M_1, Y_2^{t-1})$$

$$+ \sum_{t=1}^{n} I(M_2; Y_{2,t}|M_1, Y_2^{t-1}, Y_{1,t+1}^n)$$

$$\overset{a}{=} \sum_{t=1}^{n} I(Y_2^{t-1}, M_1; Y_{1,t}|Y_{1,t+1}^n)$$

$$+ \sum_{t=1}^{n} I(M_2; Y_{2,t}|M_1, Y_2^{t-1}, Y_{1,t+1}^n)$$

$$\overset{b}{\leq} \sum_{t=1}^{n} I(Y_2^{t-1}, Y_{1,t+1}^n, M_1; Y_{1,t})$$

$$+ \sum_{t=1}^{n} I(M_2; Y_{2,t}|M_1, Y_2^{t-1}, Y_{1,t+1}^n)$$

$$= \sum_{t=1}^{n} I(U_t, X_{1,t}; Y_{1,t}) + \sum_{t=1}^{n} I(X_{2,t}; Y_{2,t}|U_t, X_{1,t})$$

(A.9)

where (a) holds because due to the Csiszar-Korner identity the second and the third mutual information functions on the left side of (a) are equal; (b) holds because conditioning does not reduce the entropy. Then by substituting (A.9) in (A.8), we derive:

$$n(R_1 + R_2) \leq \sum_{t=1}^{n} I(U_t, X_{1,t}; Y_{1,t}) + \sum_{t=1}^{n} I(X_{2,t}; Y_{2,t}|U_t, X_{1,t})$$

$$+ n(D_{12} + D_{21}) + n(\epsilon_n^1 + \epsilon_n^2)$$

(A.10)

Also, we have:

$$n(R_1 + R_2) \leq I(M_1; Y_1^n, V_{21}^{L_d}) + I(M_2; Y_2^n, V_{12}^{L_d}) + n(\epsilon_n^1 + \epsilon_n^2)$$

$$\leq I(M_1; Y_1^n) + I(M_2; Y_1^n, Y_2^n, V_{12}^{Ld})$$
$$+ I(M_1; V_{21}^{Ld}|Y_1^n) + n(\epsilon_n^1 + \epsilon_n^2)$$
$$\overset{a}{\leq} I(M_1; Y_1^n) + I(M_2; Y_1^n, Y_2^n) + H(V_{21}^{Ld})$$
$$+ n(\epsilon_n^1 + \epsilon_n^2)$$
$$\leq I(M_1; Y_1^n) + I(M_2; Y_1^n, Y_2^n|M_1) + nD_{21}$$
$$+ n(\epsilon_n^1 + \epsilon_n^2)$$
$$= I(M_1; Y_1^n) + I(M_2; Y_2^n|M_1)$$
$$+ I(M_2; Y_1^n|Y_2^n, M_1) + nD_{21} + n(\epsilon_n^1 + \epsilon_n^2)$$
$$\overset{b}{\leq} \sum_{t=1}^n I(U_t, X_{1,t}; Y_{1,t}) + \sum_{t=1}^n I(X_{2,t}; Y_{2,t}|U_t, X_{1,t})$$
$$+ \sum_{t=1}^n I(X_{2,t}; Y_{1,t}|Y_{2,t}, X_{1,t}, U_t)$$
$$+ nD_{21} + n(\epsilon_n^1 + \epsilon_n^2)$$
$$= \sum_{t=1}^n I(U_t, X_{1,t}; Y_{1,t}) + I(X_{2,t}; Y_{1,t}, Y_{2,t}|X_{1,t}, U_t)$$
$$+ nD_{21} + n(\epsilon_n^1 + \epsilon_n^2)$$
$$= \sum_{t=1}^n I(U_t, X_{1,t}; Y_{1,t}) + \sum_{t=1}^n I(X_{2,t}; Y_{1,t}|U_t, X_{1,t})$$
$$+ \sum_{t=1}^n I(X_{2,t}; Y_{2,t}|Y_{1,t}, X_{1,t}, U_t)$$
$$+ nD_{21} + n(\epsilon_n^1 + \epsilon_n^2)$$
$$= \sum_{t=1}^n I(X_{1,t}, X_{2,t}; Y_{1,t})$$
$$+ \sum_{t=1}^n I(X_{2,t}; Y_{2,t}|Y_{1,t}, X_{1,t}, U_t)$$
$$+ nD_{21} + n(\epsilon_n^1 + \epsilon_n^2)$$
(A.11)

where the inequality (a) holds because $V_{12}^{Ld}$ is given by a deterministic function of $(Y_1^n, Y_2^n)$ and also conditioning does not reduce the entropy; the inequality (b) is derived by following the same lines as in (A.9) and (A.5). Lastly, we can derive:

$$n(R_1 + R_2) \leq I(M_1, M_2; Y_1^n, Y_2^n, V_{21}^{Ld}, V_{12}^{Ld}) + n(\epsilon_n^1 + \epsilon_n^2)$$
$$\overset{a}{=} I(M_1, M_2; Y_1^n, Y_2^n) + n(\epsilon_n^1 + \epsilon_n^2)$$
$$\leq \sum_{t=1}^n I(X_{1,t}, X_{2,t}; Y_{1,t}, Y_{2,t}) + n(\epsilon_n^1 + \epsilon_n^2)$$
(A.12)

where (a) holds because $V_{21}^{Ld}$ and $V_{12}^{Ld}$ are given by deterministic functions of $(Y_1^n, Y_2^n)$. By collecting (A.2), (A.3), (A.4), (A.7), (A.10), (A.11), (A.12) and applying a standard time-sharing argument, we derive desired constraints of (3) including those given by the auxiliary random variable $U$. The remaining constraints of (3) can be indeed derived symmetrically. The proof is thus complete. ∎

## APPENDIX II
### PROOF OF THEOREM 6

Consider a length-$n$ code with vanishing average error probability for the Gaussian IC (1) with conferencing decoders. Consider also the auxiliary random variables defined in (A.1). We need to derive the constraints in (17). Define the genie signals $G_{1,t}, G_{2,t}, \tilde{G}_{1,t}$, and $\tilde{G}_{2,t}$ as follows:

$$\begin{cases} G_{1,t} \triangleq S_{21,t} X_{1,t} + Z_{2,t} \\ G_{2,t} \triangleq S_{12,t} X_{2,t} + Z_{1,t} \\ \tilde{G}_{1,t} \triangleq S_{21,t} X_{1,t} + \tilde{Z}_{2,t} \\ \tilde{G}_{2,t} \triangleq S_{12,t} X_{2,t} + \tilde{Z}_{1,t} \end{cases} \quad t = 1, \dots, n$$
(A.13)

where $\{\tilde{Z}_{1,t}\}_{t=1}^n$ and $\{\tilde{Z}_{2,t}\}_{t=1}^n$ are zero-mean unit-variance Gaussian random processes which are independent of all other random variables. Based on Fano's inequality we have:

$$n(R_1 + R_2)$$
$$\leq I(M_1; Y_1^n, V_{21}^{Ld}) + I(M_2; Y_2^n, V_{12}^{Ld}) + n(\epsilon_n^1 + \epsilon_n^2)$$
$$\leq I(M_1; Y_1^n, G_1^n, V_{21}^{Ld}) + I(M_2; Y_2^n, G_2^n, V_{12}^{Ld}) + n(\epsilon_n^1 + \epsilon_n^2)$$
$$= I(M_1; Y_1^n, G_1^n) + I(M_2; Y_2^n, G_2^n) + I(M_1; V_{21}^{Ld}|Y_1^n, G_1^n)$$
$$+ I(M_2; V_{12}^{Ld}|Y_2^n, G_2^n) + n(\epsilon_n^1 + \epsilon_n^2)$$
$$\leq I(X_1^n; Y_1^n, G_1^n) + I(X_2^n; Y_2^n, G_2^n) + n(D_{12} + D_{21})$$
$$+ n(\epsilon_n^1 + \epsilon_n^2)$$
$$= I(X_1^n; G_1^n) + I(X_2^n; G_2^n) + I(X_1^n; Y_1^n|G_1^n)$$
$$+ I(X_2^n; Y_2^n|G_2^n) + n(D_{12} + D_{21}) + n(\epsilon_n^1 + \epsilon_n^2)$$
$$\overset{a}{=} H(G_1^n) - H(G_1^n|X_1^n) + H(Y_1^n|G_1^n) - H(Y_1^n|X_1^n)$$
$$+ H(G_2^n) - H(G_2^n|X_2^n) + H(Y_2^n|G_2^n)$$
$$- H(Y_2^n|X_2^n) + n(D_{12} + D_{21}) + n(\epsilon_n^1 + \epsilon_n^2)$$
$$= H(G_1^n) - H(Z_2^n) + H(Y_1^n|G_1^n) - H(G_2^n)$$
$$+ H(G_2^n) - H(Z_1^n) + H(Y_2^n|G_2^n)$$
$$- H(G_1^n) + n(D_{12} + D_{21}) + n(\epsilon_n^1 + \epsilon_n^2)$$
$$\overset{b}{=} H(Y_1^n|G_1^n) - H(Y_1^n|X_1^n, X_2^n) + H(Y_2^n|G_2^n)$$
$$- H(Y_2^n|X_1^n, X_2^n) + n(D_{12} + D_{21}) + n(\epsilon_n^1 + \epsilon_n^2)$$
$$= H(Y_1^n|G_1^n) - H(Y_1^n|X_1^n, X_2^n, G_1^n)$$
$$+ H(Y_2^n|G_2^n) - H(Y_2^n|X_1^n, X_2^n, G_2^n)$$
$$+ n(D_{12} + D_{21}) + n(\epsilon_n^1 + \epsilon_n^2)$$
$$= I(X_1^n, X_2^n; Y_1^n|G_1^n) + I(X_1^n, X_2^n; Y_2^n|G_2^n)$$
$$+ n(D_{12} + D_{21}) + n(\epsilon_n^1 + \epsilon_n^2)$$
$$\leq \sum_{t=1}^n I(X_{1,t}, X_{2,t}; Y_{1,t}|G_{1,t})$$
$$+ \sum_{t=1}^n I(X_{1,t}, X_{2,t}; Y_{2,t}|G_{2,t}) + n(D_{12} + D_{21})$$
$$+ n(\epsilon_n^1 + \epsilon_n^2)$$
(A.14)

where equality (a) holds because $G_i^n \to X_i^n \to Y_i^n, i = 1,2$ forms a Markov chain, and equality (b) holds because $H(Y_i^n|X_1^n, X_2^n) = H(Z_i^n), i = 1,2$. We next derive constraints on the linear combination of the rates $R_1 + 2R_2$. We can write:

$$n(R_1 + 2R_2) \leq I(M_2; Y_2^n, V_{12}^{Ld}) + I(M_1; Y_1^n, V_{21}^{Ld})$$
$$+ I(M_2; Y_2^n, V_{12}^{Ld}) + n(\epsilon_n^1 + 2\epsilon_n^2)$$
$$= I(M_2; Y_2^n) + I(M_2; V_{12}^{Ld}|Y_2^n) + I(M_1; Y_1^n)$$
$$+ I(M_1; V_{21}^{Ld}|Y_1^n) + I(M_2; Y_2^n) + I(M_2; V_{12}^{Ld}|Y_2^n)$$
$$+ n(\epsilon_n^1 + 2\epsilon_n^2)$$

$$\leq I(M_2; Y_2^n, M_1) + I(M_1; Y_1^n, G_1^n) + I(M_2; Y_2^n)$$
$$+ n(2D_{12} + D_{21}) + n(\epsilon_n^1 + 2\epsilon_n^2)$$
$$= I(M_2; Y_2^n | M_1) + I(M_1; Y_1^n | G_1^n) + I(X_1^n; G_1^n)$$
$$+ I(X_2^n; Y_2^n) + n(2D_{12} + D_{21}) + n(\epsilon_n^1 + 2\epsilon_n^2)$$
(A.15)

Then, for the first two mutual information functions on the right side of the last equality in (A.15) we have:

$$I(M_2; Y_2^n | M_1) + I(M_1; Y_1^n | G_1^n)$$
$$= H(Y_2^n | M_1) - H(Y_1^n | M_1) - H(Y_2^n | M_1, M_2) + H(Y_1^n | G_1^n)$$
$$\stackrel{a}{=} \sum_{t=1}^n H(Y_{2,t} | M_1, Y_2^{t-1}, Y_{1,t+1}^n)$$
$$- \sum_{t=1}^n H(Y_{1,t} | M_1, Y_2^{t-1}, Y_{1,t+1}^n)$$
$$- H(Y_2^n | M_1, M_2) + H(Y_1^n | G_1^n)$$
$$\stackrel{b}{\leq} \sum_{t=1}^n H(Y_{2,t} | M_1, Y_2^{t-1}, Y_{1,t+1}^n)$$
$$- \sum_{t=1}^n H(Y_{1,t} | M_1, Y_2^{t-1}, Y_{1,t+1}^n)$$
$$- \sum_{t=1}^n H(Y_{2,t} | M_1, M_2, Y_2^{t-1}, Y_{1,t+1}^n) + \sum_{t=1}^n H(Y_{1,t} | G_{1,t})$$
$$\stackrel{c}{=} \sum_{t=1}^n I(X_{2,t}; Y_{2,t} | U_t, X_{1,t}) + I(U_t, X_{1,t}; Y_{1,t} | G_{1,t})$$
(A.16)

where (a) is derived by [22, Lemma 2]; (b) holds because $Y_2^{t-1}, Y_{1,t+1}^n \to M_1, M_2 \to Y_{2,t}$ forms a Markov chain and conditioning does not reduce the entropy; (c) holds because $X_{i,t}$ is given by a deterministic function of $M_i$ and $G_{1,t} \to U_t, X_{1,t} \to Y_{1,t}$ forms a Markov chain. Now by substituting (A.16) in (A.15), we obtain:

$$n(R_1 + 2R_2)$$
$$\leq \sum_{t=1}^n I(X_{2,t}; Y_{2,t} | U_t, X_{1,t}) + I(U_t, X_{1,t}; Y_{1,t} | G_{1,t})$$
$$+ I(X_1^n; G_1^n) + I(X_2^n; Y_2^n) + n(2D_{12} + D_{21}) + n(\epsilon_n^1 + 2\epsilon_n^2)$$
$$= \sum_{t=1}^n I(X_{2,t}; Y_{2,t} | U_t, X_{1,t}) + I(U_t, X_{1,t}; Y_{1,t} | G_{1,t})$$
$$+ H(G_1^n) - H(Z_2^n) + H(Y_2^n) - H(G_1^n)$$
$$+ n(2D_{12} + D_{21}) + n(\epsilon_n^1 + 2\epsilon_n^2)$$
$$= \sum_{t=1}^n I(X_{2,t}; Y_{2,t} | U_t, X_{1,t}) + I(U_t, X_{1,t}; Y_{1,t} | G_{1,t})$$
$$+ I(X_1^n, X_2^n; Y_2^n) + n(2D_{12} + D_{21}) + n(\epsilon_n^1 + 2\epsilon_n^2)$$
$$\leq \sum_{t=1}^n I(X_{2,t}; Y_{2,t} | U_t, X_{1,t}) + I(U_t, X_{1,t}; Y_{1,t} | G_{1,t})$$
$$+ \sum_{t=1}^n I(X_{1,t}, X_{2,t}; Y_{2,t}) + n(2D_{12} + D_{21}) + n(\epsilon_n^1 + 2\epsilon_n^2)$$
(A.17)

We can also derive:

$$n(R_1 + 2R_2)$$
$$\leq I(M_2; Y_2^n, V_{12}^{Ld}) + I(M_1; Y_1^n, V_{21}^{Ld})$$
$$+ I(M_2; Y_2^n, V_{12}^{Ld}) + n(\epsilon_n^1 + 2\epsilon_n^2)$$
$$\leq I(M_2; Y_1^n, Y_2^n) + I(M_1; Y_1^n) + I(M_1; V_{21}^{Ld} | Y_1^n)$$
$$+ I(M_2; Y_2^n) + I(M_2; V_{12}^{Ld} | Y_2^n) + n(\epsilon_n^1 + 2\epsilon_n^2)$$
$$\leq I(M_2; Y_1^n, Y_2^n, M_1) + I(M_1; Y_1^n, G_1^n)$$
$$+ I(M_2; Y_2^n) + n(D_{12} + D_{21}) + n(\epsilon_n^1 + 2\epsilon_n^2)$$
$$= I(M_2; Y_1^n | Y_2^n, M_1) + I(M_2; Y_2^n | M_1) + I(M_1; Y_1^n | G_1^n)$$
$$+ I(X_1^n; G_1^n) + I(X_2^n; Y_2^n)$$
$$+ n(D_{12} + D_{21}) + n(\epsilon_n^1 + 2\epsilon_n^2)$$
$$\stackrel{a}{\leq} \sum_{t=1}^n I(X_{2,t}; Y_{1,t} | Y_{2,t}, X_{1,t}, M_1, Y_2^{t-1}, Y_{1,t+1}^n, Y_{2,t+1}^n)$$
$$+ \sum_{t=1}^n I(X_{2,t}; Y_{2,t} | U_t, X_{1,t}) + I(U_t, X_{1,t}; Y_{1,t} | G_{1,t})$$
$$+ I(X_1^n; G_1^n) + I(X_2^n; Y_2^n)$$
$$+ n(D_{12} + D_{21}) + n(\epsilon_n^1 + 2\epsilon_n^2)$$
$$= \sum_{t=1}^n I(X_{2,t}; Y_{1,t} | Y_{2,t}, X_{1,t}, U_t, Y_{2,t+1}^n)$$
$$+ \sum_{t=1}^n I(X_{2,t}; Y_{2,t} | U_t, X_{1,t}) + I(U_t, X_{1,t}; Y_{1,t} | G_{1,t})$$
$$+ H(G_1^n) - H(Z_2^n) + H(Y_2^n) - H(G_1^n)$$
$$+ n(D_{12} + D_{21}) + n(\epsilon_n^1 + 2\epsilon_n^2)$$
$$\stackrel{b}{\leq} \sum_{t=1}^n I(X_{2,t}; Y_{1,t} | Y_{2,t}, X_{1,t}, U_t)$$
$$+ \sum_{t=1}^n I(X_{2,t}; Y_{2,t} | U_t, X_{1,t}) + I(U_t, X_{1,t}; Y_{1,t} | G_{1,t})$$
$$+ \sum_{t=1}^n I(X_{1,t}, X_{2,t}; Y_{2,t})$$
$$+ n(D_{12} + D_{21}) + n(\epsilon_n^1 + 2\epsilon_n^2)$$
$$= \sum_{t=1}^n I(X_{2,t}; Y_{1,t}, Y_{2,t} | U_t, X_{1,t})$$
$$+ \sum_{t=1}^n I(U_t, X_{1,t}; Y_{1,t} | G_{1,t}) + \sum_{t=1}^n I(X_{1,t}, X_{2,t}; Y_{2,t})$$
$$+ n(D_{12} + D_{21}) + n(\epsilon_n^1 + 2\epsilon_n^2)$$
(A.18)

where (a) is due to (A.16), and (b) holds because conditioning does not reduce the entropy. Finally, we can write:

$$n(R_1 + 2R_2) \leq I(M_1, M_2; Y_1^n, Y_2^n, V_{12}^{Ld}, V_{21}^{Ld})$$
$$+ I(M_2; Y_2^n, V_{12}^{Ld}) + n(\epsilon_n^1 + 2\epsilon_n^2)$$
$$= I(M_1, M_2; Y_1^n, Y_2^n) + I(M_2; Y_2^n)$$
$$+ I(M_2; V_{12}^{Ld} | Y_2^n) + n(\epsilon_n^1 + 2\epsilon_n^2)$$
$$\leq I(X_1^n, X_2^n; Y_1^n, Y_2^n, \tilde{G}_1^n) + I(X_2^n; Y_2^n)$$
$$+ nD_{12} + n(\epsilon_n^1 + 2\epsilon_n^2)$$
$$= I(X_1^n, X_2^n; Y_1^n, Y_2^n | \tilde{G}_1^n) + I(X_1^n, X_2^n; \tilde{G}_1^n)$$
$$+ I(X_2^n; Y_2^n) + nD_{12} + n(\epsilon_n^1 + 2\epsilon_n^2)$$
$$= I(X_1^n, X_2^n; Y_1^n, Y_2^n | \tilde{G}_1^n) + H(\tilde{G}_1^n) - H(\tilde{Z}_2^n)$$
$$+ H(Y_2^n) - H(G_1^n) + nD_{12} + n(\epsilon_n^1 + 2\epsilon_n^2)$$
$$= I(X_1^n, X_2^n; Y_1^n, Y_2^n | \tilde{G}_1^n) + I(X_1^n, X_2^n; Y_2^n)$$
$$+ nD_{12} + n(\epsilon_n^1 + 2\epsilon_n^2)$$
$$\leq \sum_{t=1}^n I(X_{1,t}, X_{2,t}; Y_{1,t}, Y_{2,t} | \tilde{G}_{1,t})$$
$$+ \sum_{t=1}^n I(X_{1,t}, X_{2,t}; Y_{2,t}) + nD_{12} + n(\epsilon_n^1 + 2\epsilon_n^2)$$
(A.19)

Finally, by applying a standard time-sharing argument to (A.14), (A.17), (A.18), and (A.19), we derive $1^{st}, 3^{rd}, 5^{th}$, and $7^{th}$ constraints of (17), respectively. The remaining constraints of (17) could be symmetrically derived (similar to (A.17), (A.18), and (A.19)). The proof is thus complete. ∎